\documentclass{aa}
\usepackage{graphicx}
\usepackage{times}
\begin{document}

\def\kms{\mbox{km\,s$^{-1}$}}
\def\Hubble{\mbox{km\,s$^{-1}$\,Mpc$^{-1}$}}
\def\Doppler{\mathcal{D}}
\def\lsim{\raisebox{-.5ex}{$\;\stackrel{<}{\sim}\;$}}
\def\gsim{\raisebox{-.5ex}{$\;\stackrel{>}{\sim}\;$}}
\def\Snutspace{$(S,\nu,t)$-space}
\def\lgSnutspace{$(\lg S,\lg \nu,\lg t)$-space}
\newcommand{\mrm}[1]{\mathrm{#1}}
\newcommand{\dmrm}[1]{_{\mathrm{#1}}}
\newcommand{\umrm}[1]{^{\mathrm{#1}}}
\newcommand{\Frac}[2]{\left(\frac{#1}{#2}\right)}
\newcommand{\eqref}[1]{Eq.~(\ref{#1})}
\newcommand{\eqsref}[2]{Eqs~(\ref{#1}) and (\ref{#2})}
\newcommand{\eqssref}[2]{Eqs~(\ref{#1}) to (\ref{#2})}
\newcommand{\figref}[1]{Fig.~\ref{fig:#1}}
\newcommand{\tabref}[1]{Table~\ref{tab:#1}}
\newcommand{\secref}[1]{Sect.~\ref{sec:#1}}

\thesaurus{03(02.18.5; 11.01.2; 11.10.1; 11.17.4 3C 273; 13.18.1)}
\title{Modelling 20 years of synchrotron flaring in the jet of 3C~273}
\author{
M. T\"urler \inst{1,2} \and
T.J.-L. Courvoisier \inst{1,2} \and
S. Paltani \inst{1,2}
}
\institute{
Geneva Observatory, ch. des Maillettes 51, CH-1290 Sauverny, Switzerland \and
\textit{INTEGRAL} Science Data Centre, ch. d'\'Ecogia 16, CH-1290 Versoix, Switzerland
}
\offprints{M. T\"urler (ISDC)}
\mail{Marc.Turler@obs.unige.ch}
\date{Received 2 March 2000 / Accepted 5 July 2000}
\maketitle

\begin{abstract}

We present a phenomenological jet model which is able to reproduce well the observed variations of the submillimetre-to-radio emission of the bright quasar 3C~273 during the last 20 years.
It is a generalization of the original shock model of Marscher \& Gear (1985), which is now able to describe an accelerating or decelerating shock wave, in a curved, non-conical and non-adiabatic jet.
The model defines the properties of a synchrotron outburst which is expected to be emitted by the jet material in a small region just behind the shock front.
By a proper parameterization of the average outburst's evolution and of the peculiarities of individual outbursts, we are able to decompose simultaneously thirteen long-term light-curves of 3C~273 in a series of seventeen distinct outbursts.
It is the first time that a model is so closely confronted to the long-term multi-wavelength variability properties of a quasar.

The ability of the model to reproduce the very different shapes of the submillimetre-to-radio light curves of 3C~273 gives strong support to the shock model of Marscher \& Gear (1985).
Indirectly, it also reinforces the idea that the outbursts seen in the light-curves are physically linked to the distinct features observed to move along the jet with apparently superluminal velocities.
The more than 5000 submillimetre-to-radio observations in the different light-curves are able to constrain the physical properties of the jet.
The results suggest, for instance, that the magnetic field behind the shock front is rather turbulent.
There is also some evidence that the jet radius does not increase linearly with distance down the jet or, alternatively, that the synchrotron emitting material decelerates with distance and/or bends away from the line-of-sight.

\end{abstract}

\section{Introduction}
\label{sec:introduction}

The theory of synchrotron emission in relativistic jets was developed around 1980 mainly by Blandford \& K\"onigl (\cite{BK79}), Marscher (\cite{M80}) and K\"onigl (\cite{K81}).
In 1985, both Marscher \& Gear (\cite{MG85}, hereafter MG85) and Hughes et al. (\cite{HAA85}) proposed a model for the emission of a shock wave propagating down a simple relativistic jet.
Both groups assumed that the jet is confined to a cone of constant opening angle and that the jet flow is adiabatic.
The computer code of Hughes et al. (\cite{HAA89a}) was able to describe the low-frequency flux density and polarization variability of BL Lacertae (Hughes et al. \cite{HAA89b}).
The model of MG85 has however the advantage to include the effects of synchrotron and inverse-Compton energy losses of the electrons, which cannot be neglected at higher frequencies.
It provides a simple explanation for the distinct components in the jet observed using very long baseline interferometry (VLBI).
Multi-wavelength total flux measurements were found to be difficult to use to constrain the shock model of MG85, because the emission of all distinct features in the jet often overlap to form a nearly flat total spectrum, as illustrated by Marscher (\cite{M88}) for the quasar NRAO~140.

It is only since 1995 that the very well sampled total flux millimetre and radio light-curves of a few sources allowed to study the spectral evolution of individual synchrotron outbursts.
By subtracting a quiescent spectrum assumed to be constant, Litchfield et al. (\cite{LSR95}) and Stevens et al. (\cite{SLR95}, \cite{SLR96}, \cite{SRG98}) could follow the early evolution of single synchrotron outbursts in 3C~279, PKS~0420$-$014, 3C~345 and \object{3C~273}, respectively.
These studies gave additional support to the MG85 model, but failed to constrain it strongly, mainly because they could only follow the evolution of an outburst until the onset of the next one.

To overcome this problem, T\"urler et al. (\cite{TCP99}, hereafter Paper~I) proposed a new approach to derive the observed properties of synchrotron outbursts, which consists in a complete decomposition of long-term multi-wavelength light-curves into a series of self-similar flaring events.
Two different approaches are presented in Paper~I to describe the evolution of these events with both time and frequency.
The so called ``light-curve approach'' is model-independent and describes empirically the shape of the light-curves of individual outbursts at different frequencies.
A second approach based on three-stage shock models, like those of MG85 or Valtaoja et al. (\cite{VTU92}), describes directly the evolution of the flaring synchrotron spectrum.
Finally, the ``hybrid approach'' of T\"urler (\cite{T00}) models strictly the shape of the synchrotron spectrum, but leaves the evolution of the spectral turnover as free as possible.
The results of this third approach, based on self-similar outbursts as in Paper~I, were found to be in very good agreement with the spectral evolution expected by the shock model of MG85.

After this last test, the next step, which is presented here, is to adapt the shock model of MG85 in order to describe physically both the average evolution of the outbursts and their individual specificity.
This generalized shock model, described in \secref{model}, takes into account the effects of an accelerating or decelerating synchrotron source in a curved, non-conical and non-adiabatic jet.
This model is confronted to the observations by fitting seventeen distinct outbursts simultaneously to thirteen submillimetre-to-radio light-curves of 3C~273.
The proper parameterization to achieve this fit is described in \secref{parameterization} and the results of this light-curve decomposition are given in \secref{results}.
In \secref{discussion}, we discuss the implications of our results, both on the global properties of the inner jet and on the peculiarities of individual outbursts.
The main results of this study are summarized in \secref{summary}.

Throughout this paper, we use the convention $S_{\nu}\!\propto\!\nu^{+\alpha}$ for the spectral index $\alpha$ and we use ``$\lg$'', rather than ``$\log$'', for the decimal logarithm ``$\log_{10}$'', because of a lack of space in tables and long equations.

\section{Observational material}
\label{sec:material}

The light-curves fitted here are part of the multi-wavelength database of 3C~273 presented by T\"urler et al. (\cite{TPC99}).
The 12 light-curves from 5\,GHz to 0.35\,mm are as described in Paper~I, except that we now consider the observations up to 1999, including the most recent measurements from the Mets\"ahovi Radio Observatory in Finland and from the University of Michigan Radio Astronomy Observatory (UMRAO).

We extend to lower frequencies the analysis of Paper~I by adding a new light-curve at 2.7\,GHz.
This light-curve is constituted of observations from the Green Bank Interferometer (GBI) and from the 100\,m telescope at Effelsberg in Germany (Reich et al. \cite{RRP98}).
The fluxes obtained with the GBI at 2.7\,GHz and 2.25\,GHz include both 3C~273B (the inner jet) and 3C~273A (the hot spot at the far end of the jet), but these two components combine partially out of phase (Ghigo F.D., private communication).
As a consequence, the GBI measurements are only part of the total flux of 3C~273 and have to be multiplied by a scaling factor.

To scale the GBI measurements at 2.25\,GHz we use the contemporaneous Effelsberg single dish observations at 2.7\,GHz.
The nearly flat spectral index of 3C~273 at this frequency (T\"urler et al. \cite{TPC99}) allows us to neglect the small difference in frequency.
We obtain that the GBI fluxes at 2.25\,GHz have to be multiplied by 1.66 to fit the Effelsberg observations.
A similar calibration of the earlier GBI measurements at 2.7\,GHz is not possible due to the lack of contemporaneous single dish observations.
We therefore let the scaling factor of these fluxes as a free parameter of our fit.%
\footnote{\label{fn:gbi}We obtain a best-fit value of $1.12$ for this factor with the jet model presented in \secref{results}.}
To smooth out the dips in the GBI light-curves (cf. T\"urler et al. \cite{TPC99}), we average all GBI observations into bins of 10 days.
Finally, we do not consider the GBI measurements before 1980, to avoid a flux increase in the light-curve due to an outburst that started before 1979.
We end up with a total of 5234 observational points to constrain the shock model.

\section{Physical shock model}
\label{sec:model}

In the original shock model of MG85, the shock front was assumed to propagate with constant speed in a straight conical and adiabatic jet.
Partial generalizations of this model were already derived by Marscher (\cite{M90}) and Marscher et al. (\cite{MGT92}) in the case of a bending jet, and by Stevens et al. (\cite{SLR96}) in the two cases of a straight non-adiabatic jet and a curved adiabatic jet.
Here we further generalize the shock model of MG85 to account for the effects of an accelerating or decelerating shock front in a curved, non-conical and non-adiabatic jet.
In \secref{typical}, we describe the typical three-stage evolution of all outbursts, whereas in \secref{individual} we show how the physical conditions at the onset of the shock can influence the evolution of individual outbursts.

\begin{figure*}[tb]
\includegraphics[width=12cm]{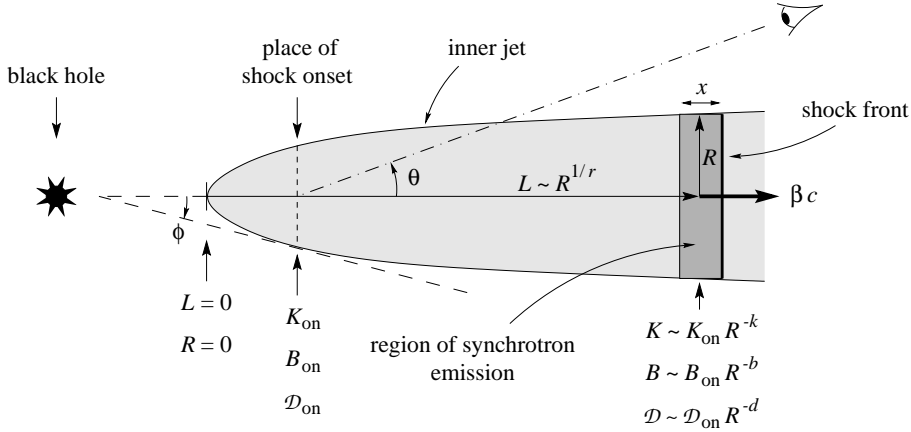}
\hfill
\parbox[b]{55mm}{
\caption{\label{fig:geometry}%
Sketch of the jet geometry discussed in \secref{model}, as observed in the rest frame of the quasar.
We assume the contribution of the inner jet to be negligible with respect to the synchrotron emission of the grey region behind the shock front.
This cylindrical portion of the jet is moving along the jet axis with a relativistic speed $\beta\,c$
}}
\end{figure*}

\subsection{Typical three-stage evolution}
\label{sec:typical}

Following MG85, we consider the synchrotron radiation emitted just behind a shock front in a cylindric portion of a jet having a radius $R$ across the jet and a length $x$ along the jet axis, as illustrated in \figref{geometry}.
Within this volume, we assume that the magnetic field $B$ is uniform in strength and nearly random in direction and that the relativistic electrons have a power law energy distribution of the form $N(E)\!=\!K\,E^{-s}$, with $N(E)dE$ being the number density of the electrons.
Measured in the rest frame of the quasar, these synchrotron emitting electrons have a relativistic bulk velocity $\beta\!=\!v/c$, with $c$ being the speed of light, and a corresponding Lorentz factor $\Gamma\!=\!(1-\beta^{2})^{-1/2}$.
As a consequence, the synchrotron emission observed with an angle $\theta$ to the jet axis is Doppler boosted with a bulk Doppler factor given by $\Doppler=\Gamma^{-1}(1-\beta\cos{\theta})^{-1}$.
We further assume that the jet opening half-angle $\phi$ is smaller than $\theta$ at any time after the onset of the shock and that $\theta$ itself is small enough to verify $\sin{(\theta+\phi)}<1/\Gamma$, so that the line-of-sight depth of the emitting region is directly proportional to its thickness $x$ (Marscher et al. \cite{MGT92}).

The observed optically thin flux density $S_{\nu}$ and turnover frequency $\nu\dmrm{m}$ of the self-absorbed synchrotron spectrum are then given by
\begin{eqnarray}
\label{S_nu}S_{\nu}&\propto&R^2\,x\,K\,B^{(s+1)/2}\,\Doppler^{(s+3)/2}\,\nu^{-(s-1)/2}\quad\quad\mbox{and}\\
\label{nu_m}\nu\dmrm{m}&\propto&\left(x\,K\,B^{(s+2)/2}\,\Doppler^{(s+2)/2}\right)^{2/(s+4)}\,,
\end{eqnarray}
where $K$ and $B$ are measured in the comoving frame of the emitting plasma and $S_{\nu}$, $\nu$, $\nu\dmrm{m}$, $R$, $x$ and $\Doppler$ are measured in the observer's frame.

The thickness $x$ of the emitting region behind the shock front is the crucial parameter that changes the outburst's evolution from one stage to the other. 
If radiative losses are the dominant cooling process of the electrons accelerated at the shock front, $x$ is approximatively given by $2\,v\dmrm{rel}\,t\dmrm{cool}$, where $v\dmrm{rel}$ is the excess velocity of the shock front relative to the emitting plasma and $t\dmrm{cool}$ is the typical cooling time of the electrons.\footnote{The centre of the emitting region is given by the typical distance $v\dmrm{rel}\,t\dmrm{cool}$ from the shock front reached by the electrons in the upstream direction before loosing substantially their energy. The full width $x$ of the emitting region is thus about twice that distance.
}
During the first stage of the shock evolution Compton scattering is the dominant cooling process of the electrons and the Compton limited thickness $x_1$ of the emitting region is given by
$x_1\!\propto\!u\dmrm{ph}^{-1}\,B^{1/2}\,\Doppler^{1/2}\,\nu^{-1/2}$, where the energy density of the synchrotron photons $u\dmrm{ph}$ can be expressed as $u\dmrm{ph}\!\propto\!K\,(B^{3s+7}\,R^{s+5})^{1/8}$ (Eq.~(13) of MG85).%
\footnote{\label{fn:order}This calculation assumes that only first order Compton scattering is important. The effect of higher order Compton losses would be to steepen even more the initial rise of the spectrum (Marscher et al. \cite{MGT92}).
}
By combining these two equations, we obtain%
\footnote{\label{fn:x}$x_1$ and $x_2$ in \eqsref{x_1}{x_2} are actually expressed in the comoving frame of the emitting plasma.
The transformation of $x_i$ to the observer's frame would be obtained by dividing these expressions by the Lorentz factor $\Gamma$ to correct for length contraction and by multiplying them by the Doppler factor $\Doppler$ to correct for front-to-back time delay (Marscher A.P., private communication).
This tricky transformation is due to the non-spherical shape of the emitting slab, which appears rotated in the observer's frame (cf. Marscher \cite{M87}; Marscher et al. \cite{MGT92}).
However, since the viewing angle $\theta$ is assumed to be smaller than about $1/\Gamma$, $\Doppler$ is proportional to $\Gamma$ (e.g. Marscher \cite{M80}), so that these two corrections cancel out when using proportionalities.
}
\begin{equation}
\label{x_1}x_1\propto R^{-(s+5)/8}\,K^{-1}\,B^{-3(s+1)/8}\,\Doppler^{1/2}\,\nu^{-1/2}\,.
\end{equation}

The shock evolution enters the synchrotron stage as soon as the photon energy density $u\dmrm{ph}$ is equal to the magnetic field energy density $u_B\!=\!B^2/(8\pi)$.
During this second stage, synchrotron radiation is the dominant energy loss process of the electrons.
Due to the similarity between Compton and synchrotron expressions of the cooling time $t\dmrm{cool}$ of the electrons, the synchrotron limited thickness $x_2$ of the emitting region is simply obtained by replacing $u\dmrm{ph}$ in the expression of $x_1$ by $u_B$, so that $x_2$ becomes
\begin{equation}
\label{x_2}x_2\propto B^{-3/2}\,\Doppler^{1/2}\,\nu^{-1/2}\,.
\end{equation}
During the final stage of the shock evolution synchrotron losses become less important than adiabatic expansion losses and the shock is assumed to remain self-similar so that $x_3\propto R$.

By substituting $x$ in \eqsref{S_nu}{nu_m} by the corresponding expression of $x_i$ $(i\!=\!1,2,3)$ for each of the three stages $i$ of the shock evolution we obtain:
\begin{eqnarray}
\label{S_nu1}S_{\nu,1}&\propto&R^{(11-s)/8}\,B^{(s+1)/8}\,\Doppler^{(s+4)/2}\,\nu^{-s/2}\,\\
\label{S_nu2}S_{\nu,2}&\propto&R^2\,K\,B^{(s-2)/2}\,\Doppler^{(s+4)/2}\,\nu^{-s/2}\\
\label{S_nu3}S_{\nu,3}&\propto&R^{3}\,K\,B^{(s+1)/2}\,\Doppler^{(s+3)/2}\,\nu^{-(s-1)/2}\\
\label{nu_m1}\nu\dmrm{m,1}&\propto&R^{-1/4}\,B^{1/4}\,\Doppler^{(s+3)/(s+5)}\\
\label{nu_m2}\nu\dmrm{m,2}&\propto&[\,K^2\,B^{s-1}\,\Doppler^{s+3}\,]^{1/(s+5)}\\
\label{nu_m3}\nu\dmrm{m,3}&\propto&[\,R\,K\,B^{(s+2)/2}\,\Doppler^{(s+2)/2}\,]^{2/(s+4)}\,.
\end{eqnarray}
Finally, by using these six last equations, it is straightforward to obtain the expressions for $S_{\mrm{m},i}\equiv S_{\nu,i}(\nu_{\mrm{m},i})$ as
\begin{eqnarray}
\label{S_m1}S\dmrm{m,1}&\propto&R^{11/8}\,B^{1/8}\,\Doppler^{(3s+10)/(s+5)}\\
\label{S_m2}S\dmrm{m,2}&\propto&R^2\,[\,K^5\,B^{2s-5}\,\Doppler^{3s+10}\,]^{1/(s+5)}\\
\label{S_m3}S\dmrm{m,3}&\propto&[\,R^{2s+13}\,K^5\,B^{2s+3}\,\Doppler^{3s+7}\,]^{1/(s+4)}\,.
\end{eqnarray}

We now assume that each of the quantities $K$, $B$ and $\Doppler$ evolves as a power-law with the radius $R$ of the jet and we parameterize this evolution as
\begin{equation}
\label{para}
K\propto R^{-k}\qquad B\propto R^{-b}\qquad \Doppler\propto R^{-d}\,.
\end{equation}
If we replace these relations into the expressions for $\nu_{\mrm{m},i}$ and $S_{\mrm{m},i}$ (\eqssref{nu_m1}{S_m3}), the turnover $(\nu\dmrm{m},S\dmrm{m})$ of the self-absorbed synchrotron spectrum will also evolve as a power-law with $R$ during each stage $i$ of the shock evolution as
\begin{equation}
\label{evo_R}
\nu_{\mrm{m},i}\propto R^{n_i}\quad \mbox{and}\quad S_{\mrm{m},i}\propto R^{f_i}\;\Rightarrow\;S_{\mrm{m},i}\propto \nu_{\mrm{m},i}^{f_i/n_i}\,,
\end{equation}
where the exponents $n_i$ and $f_i$ are given by
\begin{eqnarray}
\label{n1}n_1&\!=\!&-(b\!+\!1)/4-d\,(s\!+\!3)/(s\!+\!5)\\
\label{n2}n_2&\!=\!&-[\,2k+b\,(s\!-\!1)+d\,(s\!+\!3)\,]/(s\!+\!5)\\
\label{n3}n_3&\!=\!&-[\,2\,(k\!-\!1)+(b+d)(s\!+\!2)\,]/(s\!+\!4)\\
\label{f1}f_1&\!=\!& (11\!-\!b)/8-d\,(3s\!+\!10)/(s\!+\!5)\\
\label{f2}f_2&\!=\!& 2-\,[\,5\,k+b\,(2s\!-\!5)+d\,(3s\!+\!10)\,]/(s\!+\!5)\\
\label{f3}f_3&\!=\!& [\,2s\!+\!13-5\,k-b\,(2s\!+\!3)-d\,(3s\!+\!7)\,]/(s\!+\!4)\,.
\end{eqnarray}

Marscher (\cite{M87}) shows that even a geometrically thin source in a relativistic jet appears inhomogeneous when observed with a small angle $\theta$ between the jet axis and the line-of-sight.
The optical depth of the source will therefore depend on the frequency, which has the effect of broadening the self-absorption turnover and leads to a lower optically thick spectral index $\alpha\dmrm{thick}$ than the typical value of $+5/2$, which holds for a homogeneous synchrotron source.
To estimate the value of the spectral index below the turnover frequency $\nu\dmrm{m}$, we can cut the source into many self-similar cylindric portions of the jet having a length $l\ll x$ proportional to their radius $R$ and chosen small enough for their synchrotron emission to be homogeneous.
The emitted spectrum of each section will have a self-absorption turnover $(\nu\dmrm{m},S\dmrm{m})$ depending on $R$ according to \eqref{evo_R} with $i\!=\!3$, because in this case $l\propto R$ replaces $x$ in \eqsref{S_nu}{nu_m}.
The inhomogeneous source behind the shock front is therefore expected to have an optically thick spectral index of $\alpha\dmrm{thick}\!=\!f_3/n_3$ due to the superimposition of the homogeneous spectra of the individual sections.
However, the finite size of the emitting region should limit the frequency range over which this flatter spectral index pertains.
We therefore expect a spectral break at a frequency $\nu\dmrm{h}$ ($<\nu\dmrm{m}$), at which the spectral index $\alpha\dmrm{thick}\!=\!f_3/n_3$ returns to its homogeneous value of $+5/2$.

A high-frequency spectral break is also expected in the optically thin part of the spectrum due to a change in the electron energy distribution induced by synchrotron and/or Compton losses.
In the case of continuous injection or re-acceleration of electrons suffering radiative losses, the optically thin spectral index $\alpha\dmrm{thin}$ is expected to steepen by a value of $-1/2$ above a frequency $\nu\dmrm{b}$ (Kardashev \cite{K62}).
The break frequency $\nu\dmrm{b}$ of the spectrum in the observer's frame is related to the break energy $E\dmrm{b}$ of the electron energy distribution as $\nu\dmrm{b}\propto B\,\Doppler\,E\dmrm{b}^2$ (e.g. Marscher \cite{M80}).
For our analysis at submillimetre-to-radio frequencies, the evolution of $\nu\dmrm{b}$ with the jet radius $R$ is only relevant during the final stage of the shock evolution.
For adiabatic expansion in two dimensions, the energy $E$ of the electrons decreases with $R$ as $E\propto R^{-2/3}$ (e.g. Gear \cite{G88}).
By using this relation and \eqref{para}, we find that the evolution of the break frequency $\nu\dmrm{b}$ during the adiabatic stage is given by
\begin{equation}
\label{nu_b}
\nu\dmrm{b,3}\propto R^{n\dmrm{b}} \quad\mbox{with}\quad n\dmrm{b}=-(4/3\!+\!b\!+\!d)\,.
\end{equation}

Until now, we related the spectral turnover $(\nu\dmrm{m},S\dmrm{m})$ and the high-frequency spectral break $n\dmrm{b}$ to the radius $R$ of the emitting region.
But since we are interested in the temporal evolution of these quantities, we have to relate $R$ to the observed time $t$ after the onset of the outburst.
According to the geometry of \figref{geometry} and the basic principles of superluminal motion (e.g. Pearson \& Zensus \cite{PZ87}), we obtain
\begin{equation}
\label{t(L)}
t=\frac{(1\!+\!z)\,\sin{\theta}}{\beta\dmrm{app}\,c}\,(L\!-\!L\dmrm{on})=\frac{(1\!+\!z)}{\beta\,c\,\Gamma\,\Doppler}\,(L\!-\!L\dmrm{on})\,,
\end{equation}
where $\beta\dmrm{app}$ is the apparent transverse velocity of the source in units of $c$ and $L$ measures the distance along the jet axis in the rest frame of the quasar.
To continue working only with proportionalities, we are forced to assume that $L\dmrm{on}$ is small with respect to $L$ during the major part of the shock evolution, so that $L\!-\!L\dmrm{on}$ tends rapidly towards $L$ during, or just after, the Compton stage.
Under this assumption and by considering that $\beta\,\Gamma\,\Doppler$ is proportional to $\Doppler^2$ for $\Gamma\gg 1$ and $\theta\lsim 1/\Gamma$ (e.g. Marscher \cite{M80}), we have $t\propto \Doppler^{-2}\,L$.
Finally, by parameterizing the opening radius $R$ of the jet with the distance $L$ as $R\propto L^r$ and by remembering that $\Doppler\propto R^{-d}$ (\eqref{para}), we obtain
\begin{equation}
\label{rho}
t\propto \Doppler^{-2}\,R^{1/r}\propto R^{\rho} \quad\mbox{with}\quad \rho=(2rd+1)/r\,.
\end{equation}
With this relation, the three-stage evolution of \eqref{evo_R} can now be expressed as a function of time $t$ rather than radius $R$, as
\begin{equation}
\label{evo_t}
\left.
\begin{array}{ll}
\nu\dmrm{m,i}\propto t^{\beta_i}\quad\mbox{with}\quad\beta_i\equiv n_i/\rho\\
S\dmrm{m,i}\propto t^{\gamma_i}\quad\mbox{with}\quad\gamma_i\equiv f_i/\rho
\end{array}
\right\}
\;\Rightarrow\;\frac{\gamma_i}{\beta_i}=\frac{f_i}{n_i}\,,
\end{equation}
where $\beta_i$ and $\gamma_i$ are defined as in Paper I and depend now on the five exponents $s$, $r$, $k$, $b$ and $d$ via the \eqssref{n1}{f3} for $n_i$ and $f_i$ $(i\!=\!1,2,3)$.
 
\subsection{Specificity of individual outbursts}
\label{sec:individual}

We described in the previous section how the spectral turnover $(\nu\dmrm{m},S\dmrm{m})$ of the emitted synchrotron spectrum behaves according to the physical properties of the jet and during each of the three stages of the shock evolution.
What we describe here is how the transitions from one stage to the other are influenced by the physical quantities at the onset of the shock.
The first transition from the Compton to the synchrotron stage is characterized by the condition $u\dmrm{ph}=u_B$, which corresponds to $x_1=x_2$ and can be expressed by the proportionality
\begin{equation}
\label{x1=x2}
R_{1|2}^{-(s+5)}\propto K_{1|2}^8\,B_{1|2}^{3(s-3)}\,,
\end{equation}
where $K_{1|2}\equiv K(t_{1|2})$, $R_{1|2}\equiv R(t_{1|2})$ and $B_{1|2}\equiv B(t_{1|2})$, with the subscript $1|2$ referring to the transition from the first to the second stage of the shock evolution at a time $t_{1|2}$ after the onset of the shock.

A similar expression can be derived for the second transition from the synchrotron to the adiabatic stage by imposing that $x_2=x_3$, which corresponds to $R_{2|3}\propto B_{2|3}^{-3/2}\Doppler_{2|3}^{1/2}\nu_{2|3}^{-1/2}$, where $\nu_{2|3}\equiv \nu\dmrm{m}(t_{2|3})$, with the subscript $2|3$ referring to the transition from the second to the third stage of the shock evolution.
By replacing $\nu_{2|3}$ in the condition above by the proportionality of \eqref{nu_m2} for $\nu\dmrm{m,2}(t_{2|3})$ -- or equivalently by that of \eqref{nu_m3} for $\nu\dmrm{m,3}(t_{2|3})$ -- we obtain
\begin{equation}
\label{x2=x3}
R_{2|3}^{-(s+5)}\propto K_{2|3}\,B_{2|3}^{2s+7}\,\Doppler_{2|3}^{-1}\,.
\end{equation}

We can now further parameterize the proportionalities of \eqref{para}, as
\begin{equation}
\label{para_on}
K\propto K\dmrm{on}\,R^{-k}\qquad B\propto B\dmrm{on}\,R^{-b}\qquad \Doppler\propto \Doppler\dmrm{on}\,R^{-d}\,,
\end{equation}
where the subscript ``$\mrm{on}$'' refers to the onset of the shock.
By including these relations expressed at the particular radii $R=R_{1|2}$ and $R=R_{2|3}$ into \eqref{x1=x2} and \eqref{x2=x3}, respectively, we obtain
\begin{equation}
\label{RrRp}
R_{1|2}^{\zeta_{1|2}}\propto  K\dmrm{on}^8\,B\dmrm{on}^{3(s-3)} \quad\mbox{and}\quad R_{2|3}^{\zeta_{2|3}}\propto  K\dmrm{on}\,B\dmrm{on}^{2s+7}\,\Doppler\dmrm{on}^{-1}\,,
\end{equation}
where the exponents $\zeta_{1|2}$ and $\zeta_{2|3}$ are given by
\begin{eqnarray}
\label{xi_r}\zeta_{1|2}&\equiv& 8k+3b(s\!-\!3)-(s\!+\!5)\\
\label{xi_p}\zeta_{2|3}&\equiv& k+b(2s\!+\!7)-d-(s\!+\!5)\,.
\end{eqnarray}

The expressions of \eqref{RrRp} relate the radii $R_{1|2}$ and $R_{2|3}$ of the jet at which the two transitions $1|2$ and $2|3$ occur to the values of the physical quantities $K$, $B$ and $\Doppler$ at, or just after, the onset of the shock.
Any change of these quantities at the onset of the shock will therefore have the effect of displacing the position in the jet (via $L\propto R^{1/r}$) at which the two transitions occur.
A stronger magnetic field $B\dmrm{on}$, for instance, will have the effect to prolong the synchrotron stage, by making it start slightly further upstream in the jet and finish much further downstream for a typical value of $s\approx 2$ and comparable values of $\zeta_{1|2}$ and $\zeta_{2|3}$.

The values of $K\dmrm{on}$, $B\dmrm{on}$ and $\Doppler\dmrm{on}$, will also influence the times $t_{1|2}$ and $t_{2|3}$ after the onset of the shock at which the two transitions are observed, as well as the frequencies $\nu_{1|2}$ and $\nu_{2|3}$ and flux densities $S_{1|2}$ and $S_{2|3}$ of the self-absorption turnover $(\nu\dmrm{m},S\dmrm{m})$ at these times.
By including the proportionalities of \eqref{para_on} into \eqref{rho} for the observed time $t$ and into \eqsref{nu_m2}{S_m2} for the evolution during the synchrotron stage of turnover frequency $\nu\dmrm{m,2}$ and the corresponding flux density $S\dmrm{m,2}$, we obtain for the first transition
\begin{eqnarray}
\label{t_rp}
t_{1|2}&\propto& \Doppler\dmrm{on}^{-2}\,R_{1|2}^\rho\\
\label{nu_rp}
\nu_{1|2}&\propto& \left(\Doppler\dmrm{on}^{s+3}\,K\dmrm{on}^2\,B\dmrm{on}^{s-1}\right)^{1/(s+5)} R_{1|2}^{n_2}\\
\label{S_rp}
S_{1|2}&\propto& \left(\Doppler\dmrm{on}^{3s+10}\,K\dmrm{on}^5\,B\dmrm{on}^{2s-5}\right)^{1/(s+5)} R_{1|2}^{f_2}\,.
\end{eqnarray}
Exactly the same expressions apply also to the transition $2|3$, because both transitions are related to the synchrotron stage.%
\footnote{\label{fn:1|2|3}We would have obtained other proportionalities by using the expressions of the Compton stage for the first transition or those of the adiabatic stage for the second transition.
However, the end result of \eqsref{Dlognu}{DlogS} would have been the same, because $\nu_{1|2}\equiv \nu\dmrm{m,1}(t_{1|2})=\nu\dmrm{m,2}(t_{1|2})$, and similarly for $S_{1|2}$, $\nu_{2|3}$ and $S_{2|3}$.}

If we now substitute the expressions of \eqref{RrRp} for $R_{1|2}$ and $R_{2|3}$, we obtain the proportionalities which define the place in the \Snutspace\ (cf. \figref{evolution}) where the transitions from one stage to the other occur according to the values of $K\dmrm{on}$, $B\dmrm{on}$ and $\Doppler\dmrm{on}$.
These proportionalities can be expressed as logarithmic shifts $\Delta\lg P\!=\!\lg P-\langle\lg P\rangle$ from an average value $\langle\lg{P}\rangle$ of the parameter $P$, which stands either for $K\dmrm{on}$, $B\dmrm{on}$ and $\Doppler\dmrm{on}$ or for $S$, $\nu$ and $t$.
The set of equations relating these shifts for the first transition can be written as
\begin{eqnarray}
\label{Dlogt}
\Delta\!\lg t_{1|2}\!&\!=\!&\!
U_{t_{1|2}}\Delta\!\lg\!K\dmrm{on}\!+\!
V_{t_{1|2}}\Delta\!\lg\!B\dmrm{on}\!+\!
W_{t_{1|2}}\Delta\!\lg\!\Doppler\dmrm{on}\\
\label{Dlognu}
\Delta\!\lg \nu_{1|2}\!&\!=\!&\!
U_{\nu_{1|2}}\Delta\!\lg\!K\dmrm{on}\!+\!
V_{\nu_{1|2}}\Delta\!\lg\!B\dmrm{on}\!+\!
W_{\nu_{1|2}}\Delta\!\lg\!\Doppler\dmrm{on}\\
\label{DlogS}
\Delta\!\lg\!S_{1|2}\!&\!=\!&\!
U_{S_{1|2}}\Delta\!\lg\!K\dmrm{on}\!+\!
V_{S_{1|2}}\Delta\!\lg\!B\dmrm{on}\!+\!
W_{S_{1|2}}\Delta\!\lg\!\Doppler\dmrm{on}.
\end{eqnarray}
Similar expressions can be written for the second transition by replacing $1|2$ in \eqssref{Dlogt}{DlogS} by $2|3$.
The $2\times 9$ expressions of the parameters $U$, $V$ and $W$ obtained for both transitions are displayed in \tabref{UVW}.
There is a great symmetry among these parameters, which reflects the fact that the two equations
\begin{eqnarray}
\Delta\lg\nu_{2|3}-\Delta\lg\nu_{1|2} & = & \beta_2(\Delta\lg t_{2|3}-\Delta\lg t_{1|2}) \quad\mbox{and}\quad \\
\Delta\lg S_{2|3}-\Delta\lg S_{1|2} & = & \gamma_2(\Delta\lg t_{2|3}-\Delta\lg t_{1|2})
\end{eqnarray}
must be verified, in accordance with \eqref{evo_t}.

Finally, we can also derive the influence of $K\dmrm{on}$, $B\dmrm{on}$ and $\Doppler\dmrm{on}$ on the high-frequency spectral break $\nu\dmrm{b,2|3}\equiv \nu\dmrm{b}(t_{2|3})$, which is given by $\nu\dmrm{b,2|3}\propto \Doppler\dmrm{on}\,B\dmrm{on}\,R_{2|3}^{n\dmrm{b}}$ (cf. \eqref{nu_b}) and corresponds to the logarithmic shift
\begin{equation}
\label{Dlognu_b}
\Delta\!\lg \nu\dmrm{b,2|3}\!=\!
U_{\nu\dmrm{b}}\Delta\!\lg\!K\dmrm{on}\!+\!
V_{\nu\dmrm{b}}\Delta\!\lg\!B\dmrm{on}\!+\!
W_{\nu\dmrm{b}}\Delta\!\lg\!\Doppler\dmrm{on},
\end{equation}
with the expressions of \tabref{UVW} for $U_{\nu\dmrm{b}}$, $V_{\nu\dmrm{b}}$ and $W_{\nu\dmrm{b}}$.

\begin{table}[tb]
\addtolength{\tabcolsep}{+5pt}
\caption{\label{tab:UVW}%
Expressions of the parameters $U$, $V$ and $W$ of \eqssref{Dlogt}{DlogS} and of \eqref{Dlognu_b} for the logarithmic shifts $\Delta\lg P$ of the parameters $P$ in the first column.
These expressions characterize the effect of varying the physical quantities $K$, $B$ and $\Doppler$ at the onset of the shock.
}
\begin{flushleft}
\begin{tabular}{@{}lccc@{}}
\hline
\rule[-0.4em]{0pt}{1.4em} & \multicolumn{1}{c}{$U$} & \multicolumn{1}{c}{$V$} &\multicolumn{1}{c}{$W$} \\
\hline
\rule[-0.7em]{0pt}{1.8em} $t_{1|2}$ &
$\frac{8\rho}{\zeta_{1|2}}$ &
$\frac{3(s-3)\rho}{\zeta_{1|2}}$ &
$-2$ \\

\rule[-0.7em]{0pt}{1.8em} $t_{2|3}$ &
$\frac{\rho}{\zeta_{2|3}}$ &
$\frac{(2s+7)\rho}{\zeta_{2|3}}$ &
$-2\!-\frac{\rho}{\zeta_{2|3}}$ \\

\rule[-0.7em]{0pt}{1.8em} $\nu_{1|2}$ &
$\frac{2}{s+5}+\frac{8n_2}{\zeta_{1|2}}$ &
$\frac{s-1}{s+5}+\frac{3(s-3)n_2}{\zeta_{1|2}}$ &
$\frac{s+3}{s+5}$ \\

\rule[-0.7em]{0pt}{1.8em} $\nu_{2|3}$ &
$\frac{2}{s+5}+\frac{n_2}{\zeta_{2|3}}$ &
$\frac{s-1}{s+5}+\frac{(2s+7)n_2}{\zeta_{2|3}}$ &
$\frac{s+3}{s+5}-\frac{n_2}{\zeta_{2|3}}$ \\

\rule[-0.7em]{0pt}{1.8em} $S_{1|2}$ &
$\frac{5}{s+5}+\frac{8f_2}{\zeta_{1|2}}$ &
$\frac{2s-5}{s+5}+\frac{3(s-3)f_2}{\zeta_{1|2}}$ &
$\frac{3s+10}{s+5}$ \\

\rule[-0.7em]{0pt}{1.8em} $S_{2|3}$ &
$\frac{5}{s+5}+\frac{f_2}{\zeta_{2|3}}$ &
$\frac{2s-5}{s+5}+\frac{(2s+7)f_2}{\zeta_{2|3}}$ &
$\frac{3s+10}{s+5}-\frac{f_2}{\zeta_{2|3}}$ \\

\rule[-0.7em]{0pt}{1.8em} $\nu\dmrm{b,2|3}$ &
$\frac{n\dmrm{b}}{\zeta_{2|3}}$ &
$1+\frac{(2s+7)n\dmrm{b}}{\zeta_{2|3}}$ &
$1-\frac{n\dmrm{b}}{\zeta_{2|3}}$ \\
\hline
\end{tabular}
\end{flushleft}
\end{table}

\section{Parameterization}
\label{sec:parameterization}

In the previous section, we described in detail the physical basis of the light-curve decomposition that we present in \secref{results}.
Here, we are more concerned by the practical aspect of the decomposition and particularly by the proper definition of the 85 parameters used by the fit.

\subsection{The synchrotron spectrum and its average evolution}
\label{sec:spectrum}

According to \secref{typical}, the evolution of a shock wave in a jet follows three distinct stages to which correspond different evolutions of the self-absorption turnover of the synchrotron spectrum emitted by the plasma behind the shock front.
The evolution of the turnover $(\nu\dmrm{m},S\dmrm{m})$ during each of the three stages is fully determined through \eqref{evo_t} by only five indices, which are the three indices $k$, $b$ and $d$ defined by \eqref{para}, the index $s$ of the electron energy distribution $N(E)$ and the index $r$ of the relation $R\propto L^r$ defining how fast the jet opens with distance $L$.
In the model presented in \secref{results}, we impose the value of $d$ to be zero and leave the four other indices as free parameters of the fit.
They are constrained by the slopes $\beta_i$ and $\gamma_i$ $(i\!=\!1,2,3)$ of the outburst's evolution in the \lgSnutspace\ (cf. \figref{evolution}) and are the most interesting parameters of the light-curve decomposition, because they describe physical properties of the jet.
The two points $(t_{1|2},\nu_{1|2},S_{1|2})$ and $(t_{2|3},\nu_{2|3},S_{2|3})$ in this space at which the transitions $1|2$ and $2|3$ from one stage to the other occur are defined by only four of these six quantities, because the slopes $\beta_2$ and $\gamma_2$ can be used to determine $\nu_{2|3}$ and $S_{2|3}$ from the four other values.

Until now, we have used eight parameters, namely $s$, $r$, $k$, $b$, $\lg t_{1|2}$, $\lg t_{2|3}$, $\lg \nu_{1|2}$ and $\lg S_{1|2}$, to fully characterize the path in the \Snutspace\ followed in average by the maximum $(\nu\dmrm{m},S\dmrm{m})$ of the self-absorbed synchrotron spectrum.
The shape of this spectrum is defined by the general expression (cf. Paper~I)
\begin{equation}
\label{spec}
S_{\nu}=S\dmrm{m}\!\Frac{\nu}{\nu\dmrm{m}}^{\!\alpha\dmrm{thick}}\frac{1\!-\!\exp{(-\tau\dmrm{m}\,(\nu/\nu\dmrm{m})^{\alpha\dmrm{thin}-\alpha\dmrm{thick}})}}{1-\exp{(-\tau\dmrm{m})}}\,,
\end{equation}
where $\tau\dmrm{m}\!\approx\!\frac{3}{2}\left(\sqrt{1-\frac{8\,\alpha\dmrm{thin}}{3\,\alpha\dmrm{thick}}}-1\right)$ is a good approximation of the optical depth $\tau_{\nu}$ at the turnover frequency $\nu\dmrm{m}$ and $S\dmrm{m}$ is the real maximum of the spectrum.%
\footnote{\label{fn:Sm}Note that $S\dmrm{m}$ as defined in \secref{typical} is not the real maximum, but the extrapolation $S\dmrm{m}\umrm{\,thin}\equiv S_{\nu}\umrm{\,thin}(\nu\dmrm{m})$ down to $\nu\dmrm{m}$ of the optically thin spectrum.
The real maximum $S\dmrm{m}$ is related to $S\dmrm{m}\umrm{\,thin}$ by $S\dmrm{m}=S\dmrm{m}\umrm{\,thin}(1-\exp{(-\tau\dmrm{m})})/\tau\dmrm{m}$, but this distinction is not important for this work, which is based on proportionalities.
}
We choose to impose the value of the optically thick spectral index $\alpha\dmrm{thick}$ to $f_3/n_3$ according to the discussion in \secref{typical}.
The optically thin spectral index $\alpha\dmrm{thin}$ is also determined by the model and has the value $-s/2$ during the first two stages before flattening to $-(s\!-\!1)/2$ at the transition $2|3$ to the adiabatic stage, according to \eqssref{S_nu1}{S_nu3}.
This flattening by $\Delta\alpha\dmrm{thin}=+1/2$ is assumed to end at the time $t_{2|3}$ of the transition, but since it cannot be instantaneous we are forced to use one parameter to define the time $t\dmrm{f}$ when it starts.
Because this spectral change is expected to begin slowly before accelerating until the transition $2|3$, we choose to describe it with a logarithmic expression of the time $t$, as
\begin{equation}
\label{alpha}
\alpha\dmrm{thin}(t)=-\frac{s}{2}+\frac{1}{2}\,\frac{\lg{(t/t\dmrm{f})}}{\lg{(t_{2|3}/t\dmrm{f})}}\quad\mbox{for}\quad t\dmrm{f} \leq t \leq t_{2|3}\,.
\end{equation}

\begin{figure}[tb]
\includegraphics[width=\hsize]{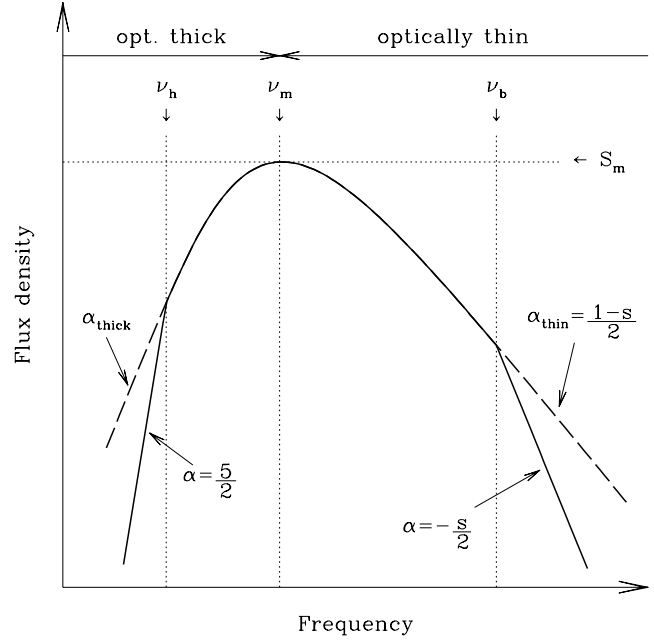}
\caption{\label{fig:synch}%
The shape of the synchrotron spectrum (solid line) assumed to be emitted behind the shock front during the final stage of the shock evolution.
We show the effect of adding two spectral breaks to the simpler spectrum (dashed line) used in Paper~I
}
\end{figure}

According to the considerations of \secref{typical}, we further allow the spectrum defined by \eqref{spec} to have a high- and a low-frequency spectral break as shown in \figref{synch}.
Since the exact shape of the breaks is difficult to assess (cf. Marscher \cite{M77}; Band \& Grindlay \cite{BG85}), we choose to keep them sharp, rather than smooth them arbitrarily, because sharp breaks have the advantage to define precisely the frequency at which they occur.
The spectrum therefore simplifies to $S\dmrm{h}\,(\nu/\nu\dmrm{h})^{2.5}$ at the lowest frequencies $(\nu\!<\!\nu\dmrm{h})$ and to $S\dmrm{b}\,(\nu/\nu\dmrm{b})^{-s/2}$ at the highest frequencies $(\nu\!>\!\nu\dmrm{b})$, where $S\dmrm{h}\equiv S_{\nu}(\nu\dmrm{h})$ and $S\dmrm{b}\equiv S_{\nu}(\nu\dmrm{b})$ are calculated by using \eqref{spec}.
The frequency ratio $\nu\dmrm{h}/\nu\dmrm{m}$ of the low-frequency break to the spectral turnover is a free parameter of the fit and is assumed to remain constant  throughout the outburst's evolution.
This ratio might actually slightly increase with time due to the increase of the thickness $x$ of the emitting region behind the shock front, but, for simplicity and because the effect might change from one stage to the other, we do not take this possible increase into account.

Since we do not extend the light-curve decomposition up to infrared frequencies, we do not need to consider a high-frequency spectral break during the two first stages of the outburst's evolution, which would have the effect to further steepen the optically thin spectral index $\alpha\dmrm{thin}$ to $-(s\!+\!1)/2$.
For frequencies above the break frequency $\nu\dmrm{b}$, $\alpha\dmrm{thin}$ remains therefore always at its value of $-s/2$, even when it is flattening at lower frequencies.
The evolution with time of the break frequency $\nu\dmrm{b}$ is given by
$\nu\dmrm{b}=\nu\dmrm{b,2|3}\,(t/t_{2|3})^{n\dmrm{b}/\rho}$, where $\nu\dmrm{b,2|3}\equiv \nu\dmrm{b}(t_{2|3})$ is a free parameter of the fit and the ratio $n\dmrm{b}/\rho$ is fixed by the values of $r$, $b$ and $d$ through \eqsref{nu_b}{rho}.

\subsection{Characteristics of different outbursts}
\label{sec:characteristics}

The $11$ parameters defined above suffice to describe fully the spectral evolution of the average outburst shown in \figref{evolution}.
The behaviour of the spectral turnover during each of the three stages of the evolution will be exactly the same for all individual outbursts.
What can change from one outburst to the other is the position in the \lgSnutspace\ where the two stage transitions occur.
In Paper~I, the two transition points were not allowed to move separately and therefore all outbursts were self-similar in the sense that they all had exactly the same evolution pattern in the \lgSnutspace.
Here, the distance between the two transition points can change and hence make the synchrotron stage relatively longer or shorter.

Although this new effect makes the light-curve modelling more complex, we do not introduce a further free parameter per outburst to describe it.
Instead, we now give a physical origin to the three logarithmic shifts $\Delta\lg S$, $\Delta\lg\nu$ and $\Delta\lg t$ in flux density $S$, frequency $\nu$ and time $t$ of Paper~I, by replacing them by shifts of the values at the onset of the shock of the factor $K$ in the expression of the electron energy distribution $N(E)$, the magnetic field $B$ and the Doppler factor $\Doppler$.
The change of the three physical quantities $K\dmrm{on}$, $B\dmrm{on}$ and $\Doppler\dmrm{on}$ results in different shifts in flux, frequency and time for the two transitions $1|2$ and $2|3$ according to \eqssref{Dlogt}{DlogS}.
The shifts used in Paper~I appear therefore to be quite unrealistic, because any such displacement in the \lgSnutspace\ will also result in making the synchrotron stage longer or shorter.

\begin{figure*}[tb]
\includegraphics[bb=20 150 590 535, width=\hsize,clip]{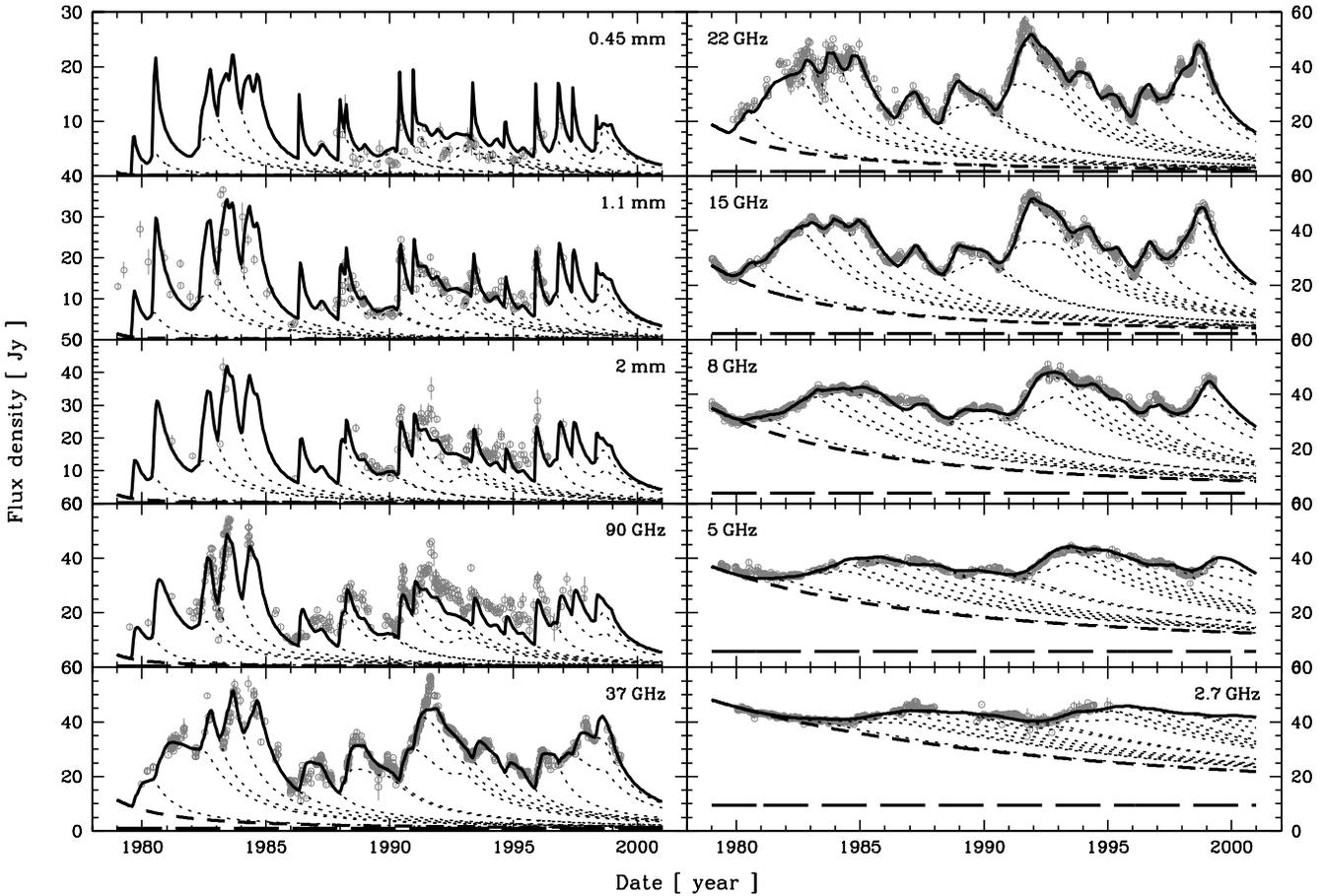}
\caption{\label{fig:lc_fit}%
Ten out of thirteen submillimetre-to-radio light-curves of 3C~273 (grey points) decomposed into a series of seventeen synchrotron outbursts.
The best-fit light-curve (solid line) is the sum of the constant emission from the outer jet (long-dashed line), the global decay of all outbursts peaking before 1979 (short-dashed line) and the seventeen superimposed outbursts (dotted lines) starting at the dates $T\dmrm{on}$ given in \tabref{shifts}
}
\end{figure*}

Actually, the physical link between the shift of an outburst and its shape is able to solve some problems we had in Paper~I (cf. \secref{results}), but does not improve the overall quality of the light-curve decomposition.
To obtain a satisfactory fit, we have to add two outbursts, one in 1988 and one in 1991.
Three other outbursts were added to extend the modelling of the light-curves to the most recent observations included here (cf. \secref{material}).
We end up with seventeen outbursts, each characterized by four parameters, namely the date of its onset $T\dmrm{on}$ and the three logarithmic changes $\Delta\lg K\dmrm{on}$, $\Delta\lg B\dmrm{on}$ and $\Delta\lg \Doppler\dmrm{on}$ of the physical quantities $K$, $B$ and $\Doppler$ at the onset of the shock.
We therefore need a total of $17\times 4=68$ parameters to model the specificity of the seventeen individual outbursts.

\subsection{Contributions from the outer jet and previous outbursts}
\label{sec:contribution}

The outer jet and especially its terminal hot spot called 3C~273A contributes significantly to the observed flux density at low radio frequencies, but this contribution can be considered as constant during the 20 years of this work.
The spectrum of the hot spot 3C~273A is shown by Conway et al. (\cite{CGP93}) and corresponds roughly to a power-law with a spectral index $\alpha$ of $0.85$ above $\sim 1$\,GHz, as described in T\"urler et al. (\cite{TPC99}).
As in Paper~I, the fit of the light-curves by the series of outbursts is done above this constant power-law contribution.

Another important contribution at low frequencies comes from the superimposed decays of the outbursts peaking before 1979.
In Paper~I, we modelled this contribution by an exponential flux decay, which was appropriate because the decline of the individual outbursts was also modelled by an exponential decay in the light-curve approach, so that the characteristic $\mrm{e}$-folding time was well constrained at all frequencies.
It would be unsatisfactory to take here the same exponential decays, because the declining phase of a three-stage evolution implies that the final light-curve decay of each outburst is a power-law and not an exponential, as illustrated by the linear ending of the logarithmic light-curves of \figref{evolution}a.

The observed flux decay $S_{\nu}^{\ast}(t)\equiv S_{\nu}(\nu^{\ast},t)$ with time $t$ at a given frequency $\nu^{\ast}$ can be written as $S_{\nu}^{\ast}(t)=S\dmrm{m}(t)(\nu^{\ast}/\nu\dmrm{m}(t))^{\alpha\dmrm{thin}}$, where the equality holds strictly if $S\dmrm{m}$ is actually the extrapolation $S\dmrm{m}\umrm{\,thin}$ of the optically thin spectrum (cf. footnote \ref{fn:Sm}).
During the final stage of the outburst's evolution, the time dependence of this quantity can be expressed as $S\dmrm{m}(t)=S\dmrm{m}^{\ast}(t/t\dmrm{m}^{\ast})^{\gamma_3}$ (cf. \eqref{evo_t}), where $t\dmrm{m}^{\ast}\equiv t\dmrm{m}(\nu^{\ast})$ is the time after the onset of the outburst when the spectral turnover $(\nu\dmrm{m},S\dmrm{m})$ passes at the frequency $\nu^{\ast}$, i.e. the time $t$ for which $\nu\dmrm{m}(t)=\nu^{\ast}$.
The corresponding equation for the turnover frequency is $\nu\dmrm{m}(t)=\nu\dmrm{m}^{\ast}(t/t\dmrm{m}^{\ast})^{\beta_3}$.
With these two equations, the light-curve decay at a given frequency $\nu^{\ast}$ during the adiabatic stage is given by
\begin{equation}
\label{S_nu(t)}
S_{\nu}^{\ast}(t)\propto \Frac{t}{t\dmrm{m}^{\ast}}^{\gamma_3-\beta_3\,\alpha\dmrm{thin}} \mbox{with}\quad t\dmrm{m}^{\ast}=t_{2|3}\Frac{\nu^{\ast}}{\nu_{2|3}}^{1/\beta_3}.
\end{equation}
The quantity $t\dmrm{m}^{\ast}$ acts as a characteristic timescale of the power-law decay at a frequency $\nu^{\ast}$ and its frequency dependence describes how this timescale increases towards lower frequencies.
In \eqref{S_nu(t)}, $\alpha\dmrm{thin}$ can be replaced by $-(s\!-\!1)/2$, if we do not consider a possible effect of the optically thin spectral break, which arises only at later times, once the break frequency $\nu\dmrm{b}$ has reached the frequency $\nu^{\ast}$.

On this basis, we can model the contribution at a given frequency $\nu$ of the superimposed decays of the outbursts peaking before 1979 by the decay of an hypothetical outburst having an amplitude $A_0(\nu)$ at the date $T_0\!=\!T\dmrm{on}\!+t\dmrm{m}(\nu)\!=\!1979.0$ and then decaying as
\begin{equation}
\label{S_nu(T)}
S_{\nu}(T)=A_0(\nu)\left(1+\frac{T\!-\!1979}{\mu\,t\dmrm{m}(\nu)}\right)^{\gamma_3+\beta_3(s-1)/2},
\end{equation}
where we have used \eqref{S_nu(t)} with $t$ replaced by $T\!-\!T\dmrm{on}=t\dmrm{m}(\nu)+T\!-\!1979$ and we have introduced a factor $\mu$ ($>1$), which is frequency independent.
There is no physical justification for introducing this factor $\mu$, but it was found to be a simple way to take into account the fact that the superimposition of several outbursts with different onset dates does decay with a longer characteristic timescale as a single outburst.
For the whole range of frequencies covered by the thirteen light-curves considered in this work, $A_0(\nu)$ in \eqref{S_nu(T)} is the millimetre-to-radio spectrum of 3C~273 in 1979.0.
This spectrum is modelled by a cubic spline parameterized at the four frequencies defined by $\lg{(\nu/\mbox{GHz})}=0.3$, $0.5$, $1.0$ and $1.5$ and is extrapolated to higher frequencies.
We choose this low frequency range, because the influence of the spectrum $A_0(\nu)$ is the greatest in the radio domain and becomes negligible at submillimetre wavelengths (cf. \figref{history_spec}).
The value and the frequency dependence of the timescale $t\dmrm{m}(\nu)$ is completely determined by the right part of \eqref{S_nu(t)} with $\nu^{\ast}$ replaced by $\nu$, so that the contribution from the outbursts prior to 1979 is only modelled by five free parameters, namely the values of $A_0(\nu)$ at the four above-mentioned frequencies and the frequency independent parameter $\mu$.

A final parameter of the fit is the GBI scaling factor mentioned in \secref{material} and therefore the total number of free parameters used to adjust the 5234 observational points in the thirteen light-curves is $11+68+6=85$.
We fit all light curves simultaneously, but usually with not more than ten of the $85$ parameters at a time.
The four indices $s$, $r$, $k$ and $b$ are adjusted together, followed by the seven other parameters defining the spectral evolution of the average outburst.
The specificity of individual outbursts is adjusted by a series of fits concerning only two or three outbursts at a time.
Finally, we fit together the six remaining parameters, which define the initial flux decay and the GBI scaling factor.
This procedure is repeated many times until the values of the parameters converge to their best-fit values.
In total, several hundreds iterative fits are needed to obtain the light-curve decomposition of \figref{lc_fit}.

\section{Results}
\label{sec:results}

We present here the results obtained by fitting the submillimetre-to-radio light-curves of 3C~273 according to the jet model described in \secref{model} and with the set of parameters defined in \secref{parameterization}.
We choose to focus in this section on the results obtained by assuming that the synchrotron emitting plasma behind the shock front moves with constant speed and direction, so that the Doppler factor $\Doppler$ remains constant during the outburst's evolution.
The index $d$ characterizing the decrease of $\Doppler$ with jet radius $R$ (cf. \eqref{para}) was therefore fixed at a value of zero.
All following figures and tables correspond to this particular jet model.
In \secref{properties}, we will however also discuss the results of an alternative jet model, in which the Doppler factor is free to vary, but we impose the jet to be conical.

The best-fit decomposition we could achieve with the jet model having a constant Doppler factor $\Doppler$ is illustrated in \figref{lc_fit}.
We note that the light-curves which are the best reproduced by the model are the radio light-curves, and especially those at 15\,GHz and at 8\,GHz.
In the 22\,GHz light-curve, the rapid variations from 1981.0 to 1983.5 are not well described by the model, and at 37\,GHz, there is mainly a problem with the prominent peak of 1991.5.
At 90\,GHz, it is striking to note that the model reproduces quite well the features of the observed light-curve, but cannot reach a sufficiently high flux level.
The same problem is also apparent at a wavelength of 2\,mm and it is only at 1.1\,mm that the model light-curve is again on average at the same flux level than the observations.
Finally, at the highest frequencies, the model light-curve does not decrease enough between the outbursts to adjust the lowest flux observations.

\begin{table}[tb]
\caption{\label{tab:param}%
Best-fit values of the parameters defined in \secref{parameterization} and corresponding to the evolution of the typical outburst shown in \figref{evolution}.
Other related quantities defined in \secref{model} are also displayed here and are distinguished from the adjusted parameters by the symbol $\dagger$
}
\begin{flushleft}
\begin{tabular}{@{}l@{~~}rl@{}rl@{~~}rl@{~~}r@{}}
\hline
\rule[-0.6em]{0pt}{1.8em}Par.& Val.& Par.& Val.& Par.& Val.& Par.& Val.\\
\hline
\rule{0pt}{1.2em}$s$& 2.05& $r$& 0.82& $k$& 3.03& $b$& 1.58\\
$t_{1|2}$& 0.07\,yr& $\nu_{1|2}$& 442\,GHz& $S_{1|2}$& 14.2\,Jy& $\zeta_{1|2}^{~~\dagger}$& 12.7\\
$t_{2|3}$& 1.06\,yr& $\nu_{2|3}^{~\dagger}$&  36.7\,GHz& $S_{2|3}^{~\dagger}$& 16.0\,Jy& $\zeta_{2|3}^{~~\dagger}$& 13.5\\
$t\dmrm{f}$& 0.75\,yr& $\nu\dmrm{b,2|3}$& 16.3\,THz& $\nu\dmrm{h}/\nu\dmrm{m}$& 0.40& $\mu$& 4.13\\
$n_1^{~\dagger}$& $-$0.64& $n_2^{~\dagger}$& $-$1.09& $n_3^{~\dagger}$& $-$1.73& $n\dmrm{b}^{~\dagger}$& $-$2.91\\
$f_1^{~\dagger}$& $+$1.18& $f_2^{~\dagger}$& $+$0.05& $f_3^{~\dagger}$& $-$1.53& $\rho^{~\dagger}$& $+$1.22\\
$\beta_1^{~\dagger}$& $-$0.53& $\beta_2^{~\dagger}$& $-$0.89& $\beta_3^{~\dagger}$& $-$1.41& & \\
$\gamma_1^{~\dagger}$& $+$0.96& $\gamma_2^{~\dagger}$& $+$0.04& $\gamma_3^{~\dagger}$& $-$1.25& & \\
$\gamma_1/\beta_1^{~\dagger}$& $-$1.83& $\gamma_2/\beta_2^{~\dagger}$& $-$0.05& $\gamma_3/\beta_3^{~\dagger}$& $+$0.88& & \\
\hline
\end{tabular}
\end{flushleft}
\end{table}

\begin{figure}[tb]
\includegraphics[bb=16 144 590 464, width=\hsize,clip]{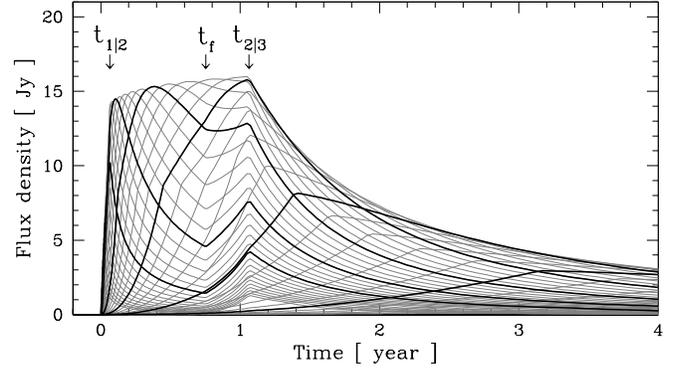}
\caption{\label{fig:evolcT}%
Model light-curves of the average outburst in 3C~273 at different frequencies spaced by 0.1\,dex (grey lines).
The six highlighted light-curves are at frequencies $\nu$ defined by $\lg{(\nu/\mbox{GHz})}\!=\!3.0,\,2.5,\ldots,\,0.5$, in order of increasing timescales
}
\end{figure}

This general discrepancy at millimetre and submillimetre frequencies is probably responsible for the relatively high reduced $\chi^2$ value of $\chi\dmrm{red}^2\!\equiv\!\chi^2/\mbox{d.o.f.}\!=\!20.5$ obtained for the overall fit as compared to the values quoted in Paper~I.
Apart from this global problem, which will be discussed in \secref{properties}, we note that the detailed structure of the light-curves is now better reproduced by the set of outbursts than it was in Paper~I.
In particular, the fast decay around 1985 is now more satisfactorily accounted for by the model.
The rapid variations of the flare peaking near 1996.0 are also better described by the new parameterization.
If we compare the high-frequency light-curves of \figref{evolcT} with those shown in Fig.~7b of Paper~I, we see that the improved description of this outburst starting in 1995.8 is achieved with a faster decline just after the peak of the high-frequency light-curves.
This faster decline results from the introduction of the new parameter $t\dmrm{f}$, which defines the time when the optically thin spectral index begins to flatten and from the fact that this time is delayed with respect to the time $t_{1|2}$ of the first stage transition.
This flattening, occurring now just at the end of the synchrotron stage, is in very good agreement with the shock model of MG85.
The above-mentioned better description of the decay around 1985 is not due to the same effect.
A satisfactory fit of 3C~273's behaviour at this epoch could indeed only be achieved once we took into account the possibility that the stage transition points $1|2$ and $2|3$ could be shifted separately as it is the case here (cf. \secref{characteristics}).
The appropriate shape of the last three outbursts peaking before 1985 was obtained by reducing very much their intermediate synchrotron stage, as shown by the values in the last column of \tabref{shifts}.

\begin{figure*}[tb]
\includegraphics[width=12cm]{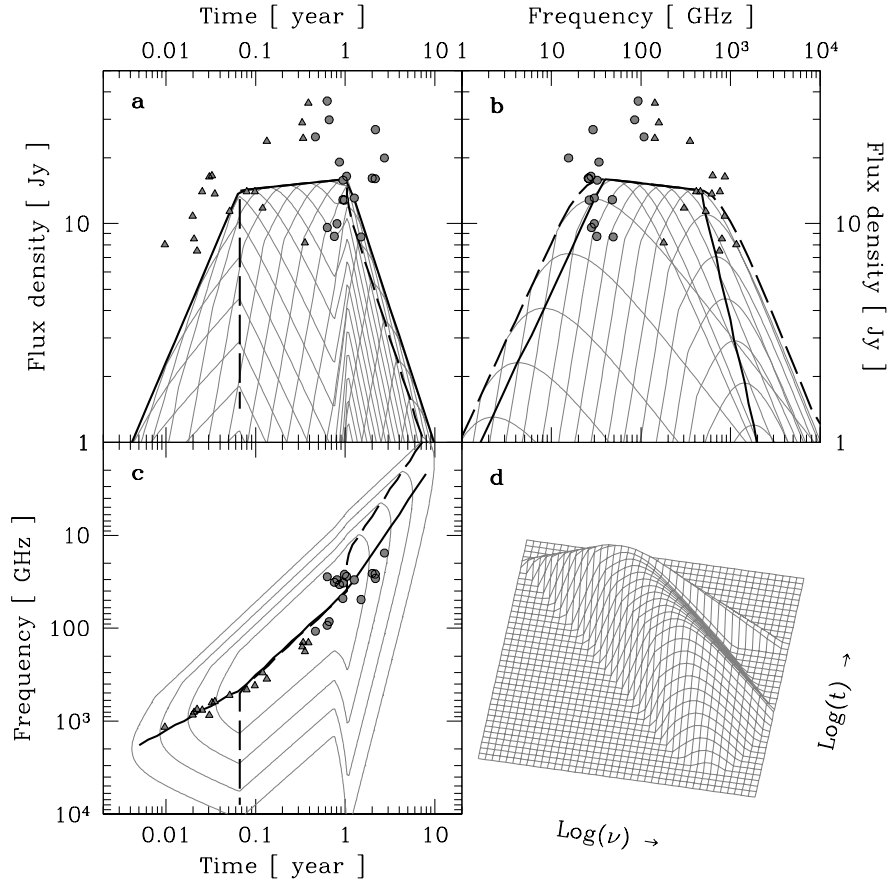}
\hfill
\parbox[b]{55mm}{
\caption{\label{fig:evolution}%
\textbf{a--d.} Logarithmic evolution of the average outburst in 3C~273.
Panel \textbf{d} shows the three-dimensional representation in the \lgSnutspace.
The other panels show the three Cartesian projections of this surface: \textbf{a} logarithmic light-curves at different frequencies spaced by 0.2\,dex; \textbf{b} synchrotron spectra at different times spaced by 0.2\,dex; \textbf{c} contour plot in the frequency versus time plane.
The thick solid line is the path followed by the turnover of the spectrum.
This line is a broken power-law having three different indices in each panel.
From left to right, these indices are $\gamma_1$, $\gamma_2$, $\gamma_3$ in panel \textbf{a}; $\gamma_3/\beta_3$, $\gamma_2/\beta_2$, $\gamma_1/\beta_1$ in panel \textbf{b}; and $\beta_1$, $\beta_2$, $\beta_3$ in panel \textbf{c}, with the values given in \tabref{param}.
The dashed line is the line connecting the peak fluxes of the light-curves at different frequencies.
The points show for each of the seventeen outbursts the position that would have the two stage transitions $1|2$ (triangles) and $2|3$ (circles) as a consequence of the logarithmic shifts $\Delta\lg K\dmrm{on}$, $\Delta\lg B\dmrm{on}$ and $\Delta\lg \Doppler\dmrm{on}$ given in \tabref{shifts}
}}
\end{figure*}

The best-fit values of the adjusted parameters and the corresponding values of other interesting quantities are displayed in \tabref{param}.
These values define the three-dimensional evolution of the average outburst in 3C~273, as shown in \figref{evolution}.
We note that the decline of the turnover flux $S\dmrm{m}$ with turnover frequency $\nu\dmrm{m}$ is now flatter than obtained with the three-stage approach of Paper~I.
Instead of $\gamma_3/\beta_3\!=\!1.14$, we now find $\gamma_3/\beta_3\!=\!0.88$.
This difference could be due to the introduction of the high-frequency break in the spectrum (cf. \figref{synch}).
Another difference is the steeper rise of $S\dmrm{m}$ with $\nu\dmrm{m}$ by nearly a factor 2 ($\gamma_1/\beta_1$ is now of $-1.83$ instead of $-0.99$).
This might be due to the introduction of the low-frequency spectral break, but it must be noticed that this initial part of the evolution is the less constrained stage of the outburst's evolution.
Its behaviour could therefore be imposed by the best-fit values of $s$, $r$, $k$ and $b$, which are more strongly constrained by the two other stages of the  evolution.
Nevertheless, both the steeper rise and the flatter decline obtained here are in better agreement with the original outburst's evolution proposed by MG85.

\begin{table}[tb]
\caption{\label{tab:shifts}%
Individual characteristics of the seventeen outbursts starting at the dates $T\dmrm{on}$ given in the first column.
The three next columns display, respectively, the logarithmic shifts of the normalization $K\dmrm{on}$ of the electron energy distribution $N(E)$, of the magnetic field $B\dmrm{on}$ and the Doppler factor $\Doppler\dmrm{on}$, with all three quantities evaluated at the onset of the shock.
The two last columns show the effect of these shifts on the relative maximum flux reached by each outburst, as expressed by $\Delta\!\lg S_{2|3}$, and the relative duration of the synchrotron stage, as expressed by $\lg (t_{2|3}/t_{1|2})$.
In \figref{evolution}a, these two last quantities characterize, respectively, the flux density level of the circles and the distance between them and the corresponding triangles
}
\begin{flushleft}
\addtolength{\tabcolsep}{-2pt}
\begin{tabular}{@{}cccccc@{}}
\hline
\rule[-0.5em]{0pt}{1.6em} $T\dmrm{on}$& $\Delta\!\lg K\dmrm{on}$& $\Delta\!\lg B\dmrm{on}$& $\Delta\!\lg \Doppler\dmrm{on}$& $\Delta\!\lg S_{2|3}$& $\lg (t_{2|3}/t_{1|2})$\\
\hline
\rule{0pt}{1.2em}$1979.56$& $+0.08$& $-0.21$& $-0.07$& $-0.09$& $0.90$\\
$1980.41$& $+0.42$& $+0.22$& $-0.02$& $+0.23$& $1.21$\\
$1982.30$& $+0.49$& $-0.58$& $-0.09$& $+0.19$& $0.14$\\
$1983.05$& $+0.67$& $-0.43$& $-0.07$& $+0.36$& $0.21$\\
$1984.02$& $+0.57$& $-0.41$& $-0.07$& $+0.27$& $0.30$\\
$1986.32$& $-0.18$& $+0.08$& $+0.06$& $-0.01$& $1.43$\\
$1987.90$& $+0.17$& $+0.19$& $-0.05$& $+0.00$& $1.34$\\
$1988.19$& $-0.48$& $+0.05$& $+0.06$& $-0.20$& $1.60$\\
$1990.34$& $-0.04$& $+0.56$& $+0.07$& $+0.10$& $1.95$\\
$1990.92$& $-0.24$& $+0.19$& $+0.09$& $+0.01$& $1.61$\\
$1991.12$& $+0.28$& $-0.32$& $-0.21$& $-0.26$& $0.63$\\
$1993.33$& $-0.39$& $+0.18$& $+0.09$& $-0.10$& $1.69$\\
$1994.64$& $-0.50$& $-0.01$& $+0.04$& $-0.26$& $1.53$\\
$1995.84$& $-0.15$& $+0.09$& $+0.08$& $+0.08$& $1.42$\\
$1996.74$& $+0.11$& $+0.21$& $-0.03$& $+0.00$& $1.40$\\
$1997.35$& $-0.11$& $+0.08$& $-0.00$& $-0.09$& $1.39$\\
$1998.30$& $-0.71$& $+0.10$& $+0.13$& $-0.22$& $1.81$\\
\hline
\end{tabular}
\end{flushleft}
\end{table}

The value of $2.05$ that we obtain for the index $s$ of the electron energy distribution $N(E)$ corresponds to an optically thin spectral index of $\alpha\dmrm{thin}\!=\!-(s-1)/2\!\approx\!-0.53$ for the final adiabatic stage of the outburst's evolution.
These values for $s$ and $\alpha\dmrm{thin}$ are slightly higher than those obtained in Paper~I, but are still well below the values of $s\!=\!2.4$ and $\alpha\dmrm{thin}\!=\!-0.7$ measured by MG85 for the strong flare of 1983.
The parameter $K$ of the electron energy distribution $N(E)$ is found to evolve with the width of the jet $R$ approximately as $K\!\propto\!R^{-3.0}$, which is just a bit steeper than the decrease $K\!\propto\!R^{-2(s+2)/3}\!\propto\!R^{-2.7}$ expected if the jet flow was adiabatic (e.g. Gear \cite{G88}).
The decrease of the magnetic field $B$ with $R$ is found to be between the two extreme cases of $B\!\propto\!R^{-2}$ expected if $\vec{B}$ was parallel to the jet axis and $B\!\propto\!R^{-1}$ expected if $\vec{B}$ was transverse to the jet (e.g. Begelman et al. \cite{BBR84}).
The obtained index of $b\!=\!1.58$ suggests that the parallel $B_{\|}$ and the perpendicular $B_{\perp}$ components of the magnetic field are roughly equal.
Finally, the best-fit value of $r\!=\!0.82$ obtained with this model having a constant Doppler factor $\Doppler$ suggests that the inner jet of 3C~273 is not really conical, but tends to open slightly less with distance $L$ along the jet axis (cf. \figref{geometry}).

The individual characteristics of the outbursts are presented in \tabref{shifts}.
They are expressed by the logarithmic shifts $\Delta\lg P\!=\!\lg P-\langle\lg P\rangle$ from an unknown average value $\langle\lg P\rangle$ of the physical quantities $K\dmrm{on}$, $B\dmrm{on}$ and $\Doppler\dmrm{on}$ at the onset of the shock.
The dispersions $\sigma$ of the seventeen values in the columns of \tabref{shifts} are $\sigma\!\approx\!0.40$ for $\Delta\lg K\dmrm{on}$; $\sigma\!\approx\!0.29$ for $\Delta\lg B\dmrm{on}$; and $\sigma\!\approx\!0.09$ for $\Delta\lg \Doppler\dmrm{on}$.
We note that the Doppler factor $\Doppler$ is the quantity with the smallest relative changes from one outburst to the other.
To study the possibility that the logarithmic shifts of the three physical quantities $K\dmrm{on}$, $B\dmrm{on}$ and $\Doppler\dmrm{on}$ are correlated, we applied a Spearman rank-order test (e.g. Press et al. \cite{PTV92}) to these three data sets.
We find out that there is a very significant anti-correlation between the values of $\Delta\lg K\dmrm{on}$ and those of $\Delta\lg \Doppler\dmrm{on}$.
The probability that a stronger anti-correlation could occur by chance is less than $6\,10^{-5}$.
A weaker trend, with a probability of non-correlation of less than 2\,\%, suggests that $\Delta\lg \Doppler\dmrm{on}$ is positively correlated with $\Delta\lg B\dmrm{on}$.
On the other hand, no significant correlation is found between $\Delta\lg K\dmrm{on}$ and $\Delta\lg B\dmrm{on}$.
These relations are discussed in \secref{peculiarities}.

\section{Discussion}
\label{sec:discussion}

In Paper~I, we discussed the relationship between the onset of an outburst and the ejection of a new VLBI component in the jet and we concluded that the identified outbursts do fairly well correspond to the VLBI knots.
The model presented here does not change this conclusion and therefore we postpone a further discussion of this relationship until more recent VLBI measurements of 3C~273 are published.
We focus, in \secref{properties}, on the global properties of the inner jet, before discussing, in \secref{peculiarities}, the peculiarities of individual outbursts.

\begin{figure*}[tb]
\includegraphics[bb=16 144 590 434,width=12cm,clip]{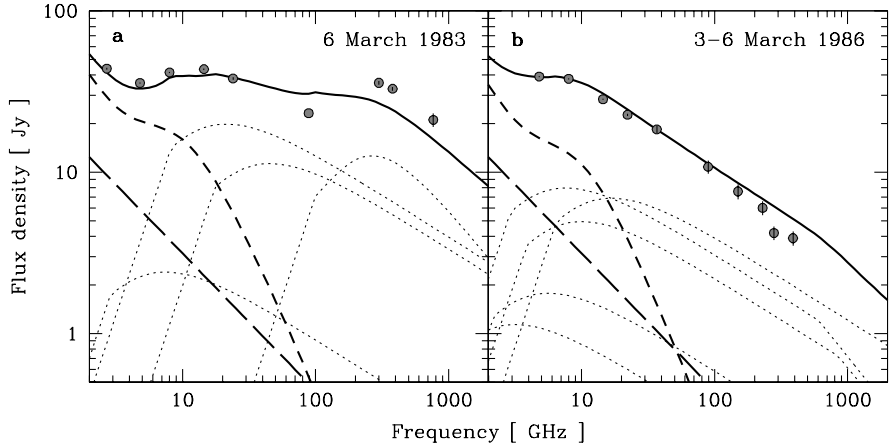}
\hfill
\parbox[b]{55mm}{
\caption{\label{fig:history_spec}%
\textbf{a and b.} Spectral dissection of 3C~273's submillimetre-to-radio emission in the 2--2000\,GHz range at the epochs of highest (\textbf{a}) and lowest (\textbf{b}) flux levels in the submillimetre range during the last 20 years.
These epochs correspond to the famous spectra of 3C~273 published by Robson et al. (\cite{RGC83}, \cite{RGB86}).
The model spectrum (solid line) is the sum of the constant outer jet contribution (long-dashed line), the slowly decaying contribution from the outbursts peaking before 1979 (short-dashed line) and the following individual synchrotron outbursts (dotted lines) evolving from high to low frequencies
}}
\end{figure*}

\subsection{On the global properties of the inner jet}
\label{sec:properties}

We have shown in \figref{lc_fit} the decomposition of a series of light-curves.
It is, of course, also possible to show the corresponding decomposition of the observed spectrum at different epochs.
As an example, we give in figure \figref{history_spec} the spectral dissection of the two most different submillimetre-to-radio spectra observed in 3C~273 during the last 20 years.
Until now, such a spectral dissection could only be achieved with VLBI observations and the best example of this is probably the decomposition by Marscher (\cite{M88}) of the radio spectrum of the relatively high redshift quasar NRAO~140.
Actually, the great similarity between the dissection of NRAO~140, based on a single epoch VLBI image, and the dissection of \figref{history_spec}, based on multi-wavelength variability observations, gives strong support to the idea that VLBI knots, observed in the jet, are physically linked to the outbursts seen in the light-curves.

The kind of spectral shape shown in \figref{history_spec}a, with a relative maximum of the spectrum in the millimetre range, was observed in several blazars by Brown et al. (\cite{BRG89}) and interpreted as the evidence of two synchrotron components contributing to the spectrum.
The two decompositions of \figref{history_spec}, suggest that these two components are not of distinct nature, but that the flaring component progressively becomes part of the more slowly varying component as it evolves towards lower frequencies.
It means that, apart from the contribution of the outer jet, the quiescent radio emission of 3C~273 can be entirely attributed to the superimposed decays of earlier outbursts.

We have assumed in this work that the emission of the inner jet is negligible outside of a small region containing the shocked plasma just behind the shock front (cf. \figref{geometry}).
It means that we have implicitly made the hypothesis that the emission of the unresolved VLBI core is due to the emission of the one or two most recent outbursts observed during the earlier phases of their evolution.
This is a rather extreme assumption, since it is usually believed that the base of the jet is also contributing to the emission, because of the higher electron and magnetic field density at the narrow end of the jet.
The present work shows that the main features of submillimetre-to-radio emission of 3C~273 can be understood as being only due to shock waves.
There are, however, some indications that there is still some place left for an underlying contribution of the inner jet.
The presence of an inner jet contribution is suggested by the relatively poor fit at millimetre and submillimetre wavelengths and especially by the fact that it is not the shape of the variations which is the greatest problem, but much more the average flux level reached by the model, which is too low at some frequencies and too high at others (cf. \secref{results}).
The idea behind this is that the model seems to make a compromise between the opposite requests of, on the one hand, the millimetre observations asking to flatten the optically thin spectral index to have more flux during the slow decline of the outbursts and, on the other hand, the submillimetre observations asking to steepen this index, in order to decrease the flux level at the end of the outbursts.

It is possible that this problem could be solved by adding an underlying nearly constant spectrum with a maximum around a frequency of 90\,GHz, in association with a steeper optically thin spectral index for the individual outbursts.
This would imply a higher value for the index $s$ of the electron energy distribution, in better agreement with the value of $2.4$ inferred from the measure of the millimetre-to-infrared slope of the flaring spectrum during the initial stage of the 1983 outburst (MG85).
Such an underlying contribution from the inner jet to the millimetre and submillimetre emission of 3C~273 could possibly be related to the so called $\mathcal{R}$ component identified in the blue-bump of 3C~273 by Paltani et al. (\cite{PCW98}).
Indeed, this slowly varying component, which dominates the optical spectrum, can be interpreted as synchrotron emission and might reveal the blazar-like characteristics of 3C~273.

As mentioned in \secref{results}, the obtained value of $1.6$ for the index $b$ suggests that the parallel $B_{\|}$ and the perpendicular $B_{\perp}$ components of the magnetic field $\vec{B}$ are roughly equal.
The simplest interpretation of this result is that the magnetic field $\vec{B}$ is turbulent behind the shock front, leading to a re-isotropization of the field.
Another possibility, but which requires much more fine tuning, is that $B_{\|}$ is dominant in the underlying jet and that the shock amplifies just enough the $B_{\perp}$ component (see MG85) to end with $B_{\perp}\!\approx\!B_{\|}$.
The obtained value of $r\simeq 0.8$ suggests that the jet opens less with distance $L$ along the jet axis than if it was conical.
Such a behaviour was recently observed in M\,87 (Virgo~A, 3C~274) on an image of its very inner jet (Junor et al. \cite{JBL99}).
A non-conical jet can arise due to an accelerating jet flow (Marscher \cite{M80}), but this interpretation would be in contradiction with the constant Doppler factor $\Doppler$ assumed in this model.
Another possibility is that there is an external pressure which confines the jet, so that the jet does not expand freely in two dimensions.
Such a collimation could be of magnetic origin, either in the case of a magnetically self-confined jet (cf. Begelman et al. \cite{BBR84}) or in the case of the two-flow model (Sol et al. \cite{SPA89}), which assumes that a highly relativistic electron-positron beam is confined within the magneto-hydrodynamic structure of a mildly relativistic electron-proton jet.

The discussion above concerns the results obtained with a shock model in which we assume a constant Doppler factor $\Doppler$ by imposing a zero value for the index $d$.
We obtain a comparable $\chi^2$ of the fit using an alternative model in which we assume that the jet is conical and thus that the index $r$ equals 1.
For this model, we obtained very similar values for most parameters given in \tabref{param}, including the indices $s$ and $b$, which therefore seem to be strongly constrained by the light-curve decomposition.
What changes with this conical jet model are the values of the indices $k$ and $d$.
The value of $k$ decreases to $2.72$, which is now almost the value of $k\dmrm{ad}=2(s\!+\!2)/3\simeq 2.70$ corresponding to an adiabatic jet flow and the best-fit value for $d$ is $0.10$, which suggests either that the emitting region is slightly decelerating while it travels down the jet or that the jet is bending away from the line-of-sight.
Even if the shock front is not decelerating, a slight deceleration of the centre of the emitting region is a natural consequence of the continuous increase of its thickness $x$ during the whole shock evolution.
Therefore, even for a straight jet, this positive value of $d$ might not be in contradiction with a value of $k$ typical for an adiabatic jet flow.
On the other hand, it is well established that the path projected on the sky followed by the VLBI components in the parsec-scale jet of 3C~273 is curved (e.g. Abraham et al. \cite{ACZ96}).
It is therefore also possible that the slight decrease of the Doppler factor $\Doppler$ suggested by the conical jet model is related to this overall bending of the jet.

The fact that both the model with a constant Doppler factor $\Doppler$ and the conical jet model give an equally good fit suggests that the light-curve decomposition cannot uniquely determine the values of the three parameters $r$, $k$ and $d$.
Indeed, by leaving all these parameters free to vary, we could not achieve a better fit than obtained with one of them fixed.
It means that besides the two models discussed above, jet models with other combinations of the values of $r$, $k$ and $d$ cannot be excluded by our light-curve decomposition.

\subsection{On the peculiarities of individual outbursts}
\label{sec:peculiarities}

Concerning the specificity of individual outbursts, we noted in Paper~I that short- and long-lived outbursts in 3C~273 are usually not peaking at the same frequency.
Long-lived outbursts were found to peak at lower frequencies and we proposed that this relationship might be related to the distance down the jet at which the shock forms.
The proper interpretation of the peculiarities of individual outbursts is now complicated by the fact that the outbursts are not self-similar anymore, because of the different behaviour of the two stage transitions $1|2$ and $2|3$ when the physical quantities $K$, $B$ and $\Doppler$ change (cf. \secref{individual}).
For instance, relatively high values of $K\dmrm{on}$ and low values of $B\dmrm{on}$ result in a shorter synchrotron stage.
This effect is particularly pronounced in the three successive outbursts between 1982 and 1984 (cf. \tabref{shifts}).
The apparent absence of a flat peaking stage in 3C~345 (Stevens et al. \cite{SLR96}) could therefore be due to a relatively low magnetic field $B$ as compared to the factor $K$ of the electron energy distribution.

We note in \secref{results} that there is apparently no correlation from one outburst to the other between the two quantities $K\dmrm{on}$ and $B\dmrm{on}$ at the onset of the shock.
This result suggests that differences from one outburst to the other are not primarily due to changes of the compression ratio $\eta$ of the shock, because in this case we would expect correlated $K\dmrm{on}$ and $B\dmrm{on}$ variations due to their similar dependence on $\eta$ (cf. MG85).
Changes of $K\dmrm{on}$ which are independent of $B\dmrm{on}$ could be possible if the acceleration process of the electrons crossing the shock front was not adiabatic.
In this case, the energy gain $\xi$ of the electrons would not simply be related to the compression ratio as $\xi=\eta^{1/3}$ (MG85), but might have different behaviours from one outburst to the other.
Since the normalization $K\dmrm{on}$ of the electron energy distribution is proportional to $\xi^{s-1}$ (MG85), changes of the electron energy gain $\xi$ for different shocks in 3C~273 could lead to the observed changes of $K\dmrm{on}$.
This interpretation is supported by the significant anti-correlation found between $K\dmrm{on}$ and the Doppler factor $\Doppler\dmrm{on}$, because a non-adiabatic shock wave converts bulk kinetic energy into internal energy (e.g. Begelman et al. \cite{BBR84}).
Outbursts with a small value of $\Doppler$ and a great value of $K$, would therefore be associated with more efficient shocks, for which a greater fraction of their bulk kinetic energy, as measured by $\Doppler$, would be converted into internal energy by increasing the average energy gain $\xi$ of the electrons.

\section{Summary and conclusion}
\label{sec:summary}

This work presents a phenomenological model describing the evolution of synchrotron outbursts expected to be emitted by shock waves in a relativistic jet.
This model, which is a generalization of the original shock model of MG85, is then confronted to the very well sampled long-term submillimetre-to-radio light-curves of 3C~273.
Many iterative fits are needed to adjust the more than 5000 observations with a series of seventeen successive outbursts defined by 85 parameters of the model.
This fitting procedure allows us not only to define the average outburst's evolution in 3C~273, but also to study the peculiarities of each individual outburst during the last 20 years.
The main results are the following:
\begin{itemize}
\item The quiescent low frequency emission can be understood as being entirely due to the superimposition of slowly decaying outbursts, which flared a few years before, and a constant contribution from the hot spot (3C~273A) of the outer jet.
\item The values of the indices describing how the physical quantities characterizing the jet evolve with its opening are found to be in general agreement with the simple jet model considered by MG85.
\item The best-fit values of these indices suggest however either that the jet opens slightly less than a conical jet if we assume that the emitting region has a constant bulk Doppler factor or, alternatively, that this Doppler factor decreases if we impose the jet to be conical.
\item In both cases, the magnetic field in the emitting region behind the shock front seems to be rather turbulent and the jet flow is found to be nearly adiabatic, especially for the conical jet model.
\item The peculiarities of individual outbursts can be understood as being due to changes at the onset of the shock of the magnetic field strength $B$, the normalization $K$ of the electron energy distribution and the Doppler factor $\Doppler$.
\item Shocks with a high value of $K$ have usually a lower value of $\Doppler$.
This anti-correlation might be related to the shock efficiency to convert bulk kinetic energy into internal energy of the plasma behind the shock front.
\end{itemize}

In the past, shock models were found to be difficult to test and constrain with total flux measurements.
We show here that very strong observational constraints can be derived from long-term multi-wavelength monitoring campaigns.
This new ability, together with interferometric imaging techniques and numerical simulations of shocks in relativistic jets (e.g. G\'omez et al. \cite{GMM97}) should lead in a near future to important progresses in the understanding of the physics involved in relativistic jets.

\begin{acknowledgements}
We thank H. Ter\"asranta for providing us previously unpublished measurements of 3C~273 at 22 and 37\,GHz from the Mets\"ahovi Radio Observatory.
This research has made use of data from the University of Michigan Radio Astronomy Observatory which is supported by funds from the University of Michigan.
\end{acknowledgements}


\begin{thebibliography}{}
\bibitem[1996]{ACZ96} Abraham Z., Carrara E.A., Zensus J.A., Unwin S.C., 1996, A\&AS 115, 543
\bibitem[1985]{BG85} Band D.L., Grindlay J.E., 1985, ApJ 298, 128
\bibitem[1984]{BBR84} Begelman M.C., Blandford R.D., Rees M.J., 1984, Rev. Mod. Phys. 56, 255
\bibitem[1979]{BK79} Blandford R.D., K\"onigl A., 1979, ApJ 232, 34
\bibitem[1989]{BRG89} Brown L.M.J., Robson E.I., Gear W.K., et al., 1989, ApJ 340, 129
\bibitem[1993]{CGP93} Conway R.G., Garrington S.T., Perley R.A., Biretta J.A., 1993, A\&A 267, 347
\bibitem[1988]{G88} Gear W.K., 1988. In: Millimetre and Submillimetre Astronomy, Wolstencroft R.D., Burton W.B. (eds.), Kluwer Academic Publishers, p. 307
\bibitem[1997]{GMM97} G\'omez J.L., Mart\'{\i} J.M., Marscher A.P., Ib\'a\~nez J.M., Alberdi A., 1997, ApJ 482, L33
\bibitem[1985]{HAA85} Hughes P.A., Aller H.D., Aller M.F., 1985, ApJ 298, 301
\bibitem[1989a]{HAA89a} Hughes P.A., Aller H.D., Aller M.F., 1989a, ApJ 341, 54
\bibitem[1989b]{HAA89b} Hughes P.A., Aller H.D., Aller M.F., 1989b, ApJ 341, 68
\bibitem[1999]{JBL99} Junor W., Biretta J.A., Livio M., 1999, Nat 401, 891
\bibitem[1962]{K62} Kardashev N.S., 1962, AZh 39, 393 (SvA 6, 317)
\bibitem[1981]{K81} K\"onigl A., 1981, ApJ 243, 700
\bibitem[1995]{LSR95} Litchfield S.J., Stevens J.A., Robson E.I., Gear W.K., 1995, MNRAS 274, 221
\bibitem[1977]{M77} Marscher A.P., 1977, ApJ 216, 244
\bibitem[1980]{M80} Marscher A.P., 1980, ApJ 235, 386
\bibitem[1987]{M87} Marscher A.P., 1987. In: Superluminal Radio Sources, Zensus J.A., Pearson T.J. (eds.), Cambridge Univ. Press, p.~280
\bibitem[1988]{M88} Marscher A.P., 1988, ApJ 334, 552
\bibitem[1990]{M90} Marscher A.P., 1990. In: Parsec-Scale Radio Jets, Zensus J.A., Pearson T.J. (eds.), Cambridge Univ. Press, p. 236
\bibitem[1985]{MG85} Marscher A.P., Gear W.K., 1985, ApJ 298, 114 (MG85)
\bibitem[1992]{MGT92} Marscher A.P., Gear W.K., Travis J.P., 1992. In: Variability of Blazars, Valtaoja E., Valtonen M. (eds.), Cambridge Univ. Press, p. 85
\bibitem[1998]{PCW98} Paltani S., Courvoisier T.J.-L., Walter R., 1998, A\&A 340, 47
\bibitem[1987]{PZ87} Pearson T.J., Zensus J.A., 1987. In: Superluminal Radio Sources, Zensus J.A., Pearson T.J. (eds.), Cambridge Univ. Press, p.~1
\bibitem[1992]{PTV92} Press W.H., Teukolsky S.A., Vetterling W.T., Flannery B.P., 1992, Numerical Recipes in FORTRAN, 2$\umrm{nd}$ ed., Cambridge Univ. Press
\bibitem[1998]{RRP98} Reich W., Reich P., Pohl M., Kothes R., Schlickeiser R., 1998, A\&AS 131, 11
\bibitem[1983]{RGC83} Robson E.I., Gear W.K., Clegg P.E., et al., 1983, Nat 305, 194
\bibitem[1986]{RGB86} Robson E.I., Gear W.K., Brown L.M.J., et al., 1986, Nat 323, 134
\bibitem[1989]{SPA89} Sol H., Pelletier G., Ass\'eo E., 1989, MNRAS 237, 411
\bibitem[1995]{SLR95} Stevens J.A., Litchfield S.J., Robson E.I., et al., 1995, MNRAS 275, 1146
\bibitem[1996]{SLR96} Stevens J.A., Litchfield S.J., Robson E.I., et al., 1996, ApJ 466, 158
\bibitem[1998]{SRG98} Stevens J.A., Robson E.I., Gear W.K., et al., 1998, ApJ 502, 182
\bibitem[2000]{T00} T\"urler M., 2000. In: Black Holes in Binaries and Galactic Nuclei, Kaper L., van den Heuvel E.P.J., Woudt P.A. (eds.), ESO Workshop, Springer-Verlag, in press
\bibitem[1999a]{TPC99} T\"urler M., Paltani S., Courvoisier T.J.-L., et al., 1999a, A\&AS 134, 89
\bibitem[1999b]{TCP99} T\"urler M., Courvoisier T.J.-L., Paltani S., 1999b, A\&A 349, 45 (Paper~I)
\bibitem[1992]{VTU92} Valtaoja E., Ter\"asranta H., Urpo S., et al., 1992, A\&A 254, 71
\end{thebibliography}
\end{document}